\documentclass[12pt]{iopart}

\usepackage{cite}
\usepackage{graphicx}

\begin{document}

\title[]{Subwavelength internal imaging by means of the wire medium}

\author{Yan Zhao, Pavel Belov and Yang Hao}

\address{School of Electronic Engineering and Computer Science, Queen Mary, University of London, Mile End Road, London, E1 4NS, United Kingdom}
\ead{yan.zhao@elec.qmul.ac.uk}

\begin{abstract}
The phenomenon of evanescent wave amplification inside of a half-wavelength-thick wire medium slab is studied numerically and used for subwavelength imaging of objects at significant distances from the lens in this paper. The wire medium is analyzed using both a spatially dispersive finite-difference time-domain (FDTD) method employing the effective medium theory and a full-wave commercial electromagnetic simulator CST Microwave Studio$^{\rm TM}$ modelling the physical structure. It is demonstrated that subwavelength details of a source placed at a distance of $\lambda/10$ from a wire medium slab can be detected inside the slab with a resolution of approximately $\lambda/10$ in spite of the fact that they cannot be resolved at the front interface of the device, due to the rapid decay of evanescent spatial harmonics in free space. The influence of different distances between the source and the wire medium on the internal imaging property is also addressed.
\end{abstract}

\maketitle

\section{Introduction}
Conventional imaging systems are restricted by the so-called diffraction limit: any source details below the half-wavelength at the frequency of operation cannot be resolved at the image plane. However, it has been proposed in \cite{Pendry} that a planar lens formed by the left-handed material (LHM) \cite{Veselago} can be used to image source information with a spatial resolution below the diffraction limit. A lens formed by the LHM with $\varepsilon=\mu=-1$ is termed as the `perfect lens' and the principle of its operation is based on the negative refraction of propagating waves and the amplification of evanescent field components \cite{Pendry}. It is well known that evanescent waves carry subwavelength source details and decay exponentially in positive-index materials. The uniqueness of the LHM is that the evanescent waves show a growing behavior when propagating within the medium. Hence the subwavelength details which are lost in the region between the source and a perfect lens can be restored to create a perfect image \cite{Pendry}. The amplification of evanescent waves in the LHM is due to the resonant excitation of surface plasmons at the interfaces. However, such effect is sensitive to the losses in the LHM thus limits the maximum thickness of the LHM slab \cite{Podolskiy}. Furthermore, the mismatch of the LHM with its surrounding medium also limits the imaging capability of LHM lenses \cite{Smith}.

Recently it has been suggested to use an alternative way to transport subwavelength source details to an image plane at a significant distance. The principle of operation, referred as `canalisation' \cite{BelovCanal}, is based on that for a certain type of devices, the evanescent wave components can be transformed into propagating waves, and therefore the source field can be delivered to its back interface with little or no deterioration. In contrast to the case of the LHM, such devices are less sensitive to losses. Recent works on the subwavelength imaging using anisotropic materials (operating in the canalisation regime) include \cite{BelovLayered,Wood,Scalora}. Moreover, the work presented in \cite{Liu} shows that introducing nonlinearity is able to suppress the diffraction, and is useful for future subwavelength imaging devices even when the dispersion curve is not flat.

One typical example of the structures operating in the canalisation regime is the wire medium formed by an array of parallel conducting wires \cite{BelovWire}. The thickness of the wire medium needs to be equal to an integer multiple of half-wavelengths at the operating frequency (due to the Fabry-Perot resonance) in order to avoid reflections between the source and the structure. It has been demonstrated experimentally and numerically that such a canalisation regime does indeed exist and, subwavelength details can be transported through the wire medium.

In addition to the canalisation regime, another interesting phenomenon of the wire medium is observed from numerical simulations and presented in this paper: the evanescent waves can be amplified \textit{inside} the wire medium. Hence the effect of images appearing inside of the wire medium slab is called `internal imaging'. The previously investigated subwavelength imaging property of the wire medium, i.e. images appear at the back interface, is highly dependent on the distance between the source and the wire medium, and the external images disappear when the distance increases. Due to the internal imaging effect, the subwavelength imaging capability of the wire medium can be further improved. In contrast to the case of LHM lenses where the amplification of evanescent waves is due to the resonant excitation of surface plasmons, the amplification of evanescent waves in a wire medium slab is a result of the resonant excitation of standing waves inside the slab. Although the internal imaging effect also exists in LHM slabs (there are two focal points at both inside and outside of LHM lenses), as mentioned above, the practical realization of LHM lenses has many difficulties. On the other hand, the internal imaging capability of the wire medium can be used for subwavelength imaging of objects at significant distances from the lens. In the following, we first present the numerical simulation results using a spatially dispersive finite-difference time-domain (FDTD) method developed in \cite{ZhaoOE}, where the effective medium theory is used and micro-structures of the wire medium are neglected. Hence the FDTD method has better efficiency comparing with simulations modelling micro-structures. Then the FDTD simulation results are validated by a commercial simulation package CST Microwave Studio$^{\rm{TM}}$ modelling the actual physical structure of the wire medium.

\section{Spatially dispersive FDTD simulations of internal imaging by the wire medium}
The simulation of the wire medium can be performed either by modelling the physical structure i.e. parallel conducting wires, or using the effective medium theory, so that the wire medium can be represented by a homogeneous dielectric slab provided that the inner spacing of wires is much smaller than the wavelength of interest. In this section, we demonstrate the internal imaging phenomenon numerically using the effective medium theory by applying the spatially dispersive FDTD method \cite{ZhaoOE}. modelling the physical structure based on a commercial simulation package CST Microwave Studio$^{\rm{TM}}$ is introduced in the next section.

The wire medium can be approximated as a homogeneous dielectric slab with both frequency and spatial dispersions. In this paper we use the following expression for the permittivity tensor derived in \cite{BelovSpatial}:
\begin{equation}
\varepsilon(\omega,\textbf{q})=\textbf{xx}+\varepsilon(\omega,q_y)\textbf{yy}+\textbf{zz},\qquad\varepsilon(\omega,q_y)=1-\frac{k_p^2}{k^2-q_y^2},
\label{eq_eff}
\end{equation}
where the $y$-axis is oriented along wires, $q_y$ is the $y$-component of the wave vector $\textbf{q}$, $k=\omega/c$ is the wave number of the free space, $c$ is the speed of light, and $k_p$ is the wave number corresponding to the plasma frequency of the wire medium. The wave number $k_p$ depends on the lattice periods $a$ and $b$, and on the radius of wires $r$ \cite{WMJEWA}:
\begin{equation}
k_p^2=\frac{2\pi/(ab)}{\ln\frac{\sqrt{ab}}{2\pi r}+F(a/b)}, \quad
F(\xi)= -\frac{1}{2}\ln\xi+ \sum\limits_{n=1}^{+\infty}
\left(\displaystyle \frac{\mbox{coth}(\pi n
\xi)-1}{n}\right)+\frac{\pi}{6}\xi.
\label{eq_k0}
\end{equation}
For the case of the square grid ($a=b$), $F(1)=0.5275$ and the expression (\ref{eq_k0}) reduces to
\begin{equation}
k_p^2=\frac{2\pi/a^2}{\ln\frac{a}{2\pi r}+0.5275}.
\label{eq_k0sq}
\end{equation}
The expression (\ref{eq_eff}) for the permittivity tensor of the wire medium is valid, and the structure is non-magnetic if the wires are thin as compared to the lattice periods (when the polarization across the wires is negligibly small as compared to the longitudinal polarization) and if the lattice periods are much smaller than the wavelength (when the wire medium can be homogenized).

A spatially dispersive FDTD method has been developed \cite{ZhaoOE} to take into account the spatial dispersion effect in the wire medium \cite{BelovSpatial}. We have applied the spatially dispersive FDTD method to analyze the subwavelength imaging property of the wire medium \cite{ZhaoOE}. In this paper, we extend the method to three dimensions to study the internal imaging property of the wire medium.

\begin{figure}[htbp]
\centering
\includegraphics[width=13cm]{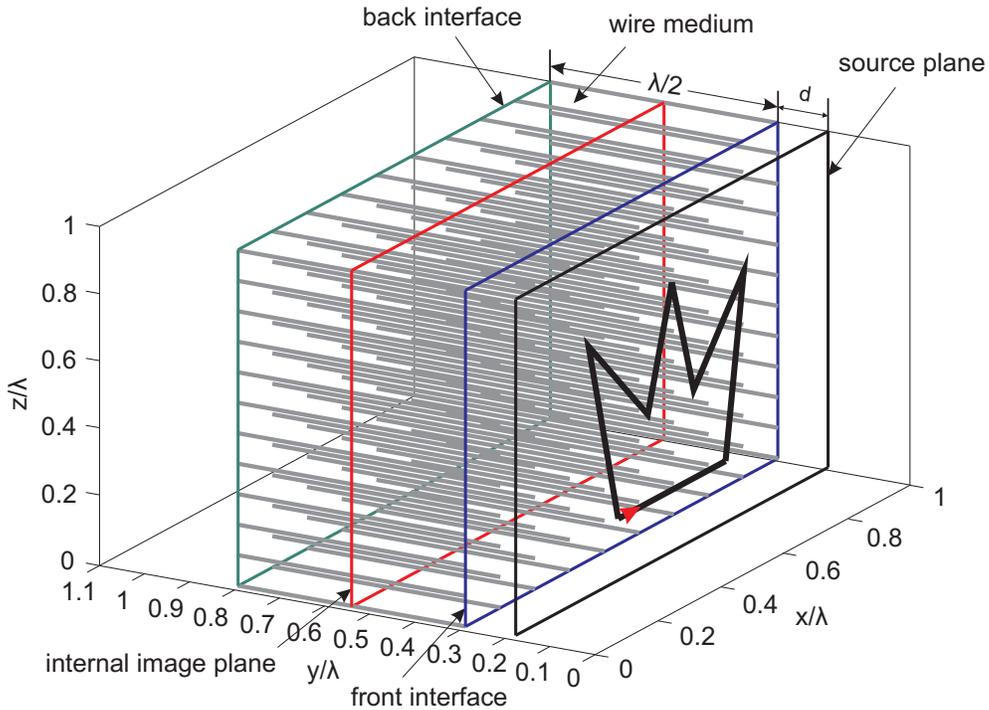}
\caption{A three-dimensional (3-D) FDTD simulation domain for the modelling of internal subwavelength imaging by a wire medium slab with thickness $\lambda/2$ and infinite transverse dimensions.}
\label{fig_domain}
\end{figure}

\begin{figure}[htbp]
\centering
\includegraphics[width=15cm]{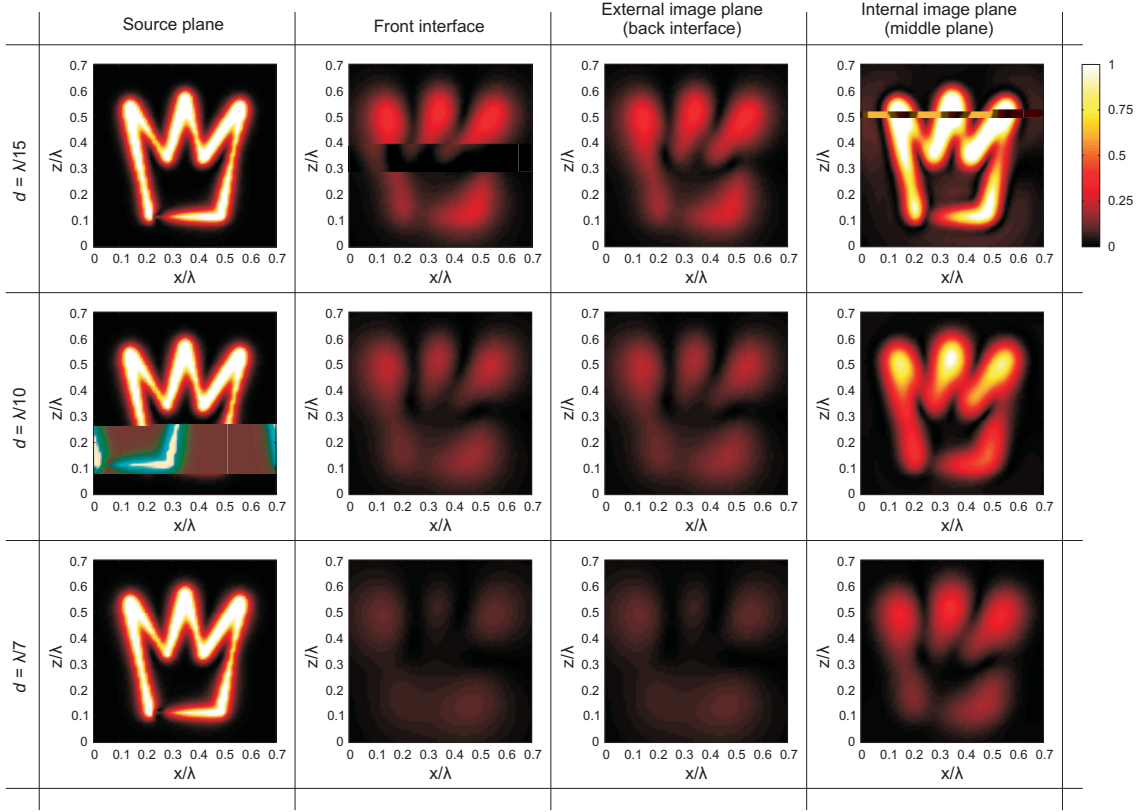}
\caption{Normalized distributions of the $y$-component of the electric field at the source plane, the front interface, and the electric displacement at the internal image plane of the wire medium for different distances between the source and the wire medium calculated from spatially dispersive FDTD simulations.}
\label{fig_fdtd}
\end{figure}

The computation domain in FDTD is shown in Fig.~\ref{fig_domain} (except that the wire medium is modelled as a bulk material instead of the parallel thin wires).
A crown-shaped near-field source is applied in the simulations and it is placed at a distance of $d$ to the front interface of the wire medium. The distance $d$ varies from $\lambda/15$, $\lambda/10$ to $\lambda/7$. The source is modelled as segments of perfect electric conductor (PEC) using staircase approximations, and it is excited at the bottom left corner using a delta-gap excitation in FDTD \cite{Taflove}. The radius of the PEC wires is the same as the FDTD cell size, which is fixed at $\lambda/100$ for all FDTD simulations. The time step is chosen according to the Courant stability criterion \cite{Taflove}. Berenger's perfectly matched layer (PML) \cite{Berenger} is employed to truncate the outer boundary of the simulation domain along $y$-direction and the modified PML \cite{ZhaoAP} is applied to model infinite wire medium slabs along $x$- and $z$-directions and reduce the convergence time in simulations \cite{ZhaoAP}. The dimensions of the simulation domain in $x$- and $z$-directions are $2\lambda/3\times2\lambda/3$. The thickness of the wire medium slab is chosen as $\lambda/2$ to achieve good impedance matching between the source and the structure. As shown in Fig.~\ref{fig_domain}, the middle plane of the wire medium is defined as the internal image plane. It has been demonstrated that the wire medium can image the electric field component along the wires \cite{BelovWire}. Inside the wire medium, the field distribution has TEM polarization (transverse electric and magnetic modes with respect to the orientation of wires) and it is formed by the so-called transmission-line modes \cite{BelovCanal}. The electric field components of such modes are zero along the wires (but the electric displacement is non-zero). Therefore, in Fig.~\ref{fig_fdtd} we plot the distributions of electric field only at the source plane and the front interface, but at the internal image plane the electric displacement (normalized to the free space permittivity) is plotted instead. It is equivalent to the electric current density flowing along the wire since $D_y=\varepsilon_0E_y+J_y/(j\omega)$. The field distributions at various planes are calculated according to different distances $d$ between the source and the wire medium and plotted after the steady state is reached in simulations.

It is clearly seen that the crown-shaped source creates a subwavelength field distribution at the source plane, however, it becomes blurred at the front interface of the wire medium due to the spatial decay of the evanescent spatial harmonics in the free space region, especially for the case of $d=\lambda/7$ that the distribution at the front interface can be hardly identified. The wire medium operates in the canalisation regime and the electric field distribution at the back interface is repeated at the front one, however, such field distribution does not create a sharp image of the original source. The surprising result is that inside the wire medium, lost subwavelength details can be recovered due to the amplification of evanescent wave components. In contrast to the case of LHM slabs, all incident evanescent spatial harmonics inside the wire medium are transformed into propagating waves, independent on their transverse variations. The thickness of the wire medium slab is tuned to the Fabry-Perot resonance. As a result, the evanescent waves excite standing waves inside the slab and in this case, they are amplified due to the natural resonance of the wire medium slab. It can be identified from Fig.~\ref{fig_fdtd} that the resolution of the internal images is about $\lambda/10$.

The internal images with good subwavelength resolutions are observed when the source is placed close to the wire medium slab (i.e. $d<\lambda/15$). If the distance is increased to $\lambda/10$ and even $\lambda/7$, the internal images do not contain enough subwavelength details of the source. We conclude that internal imaging with good subwavelength resolution is available only for a limited range of distances between the source and the wire medium, because the dependence of the transfer function for the evanescent spatial harmonics on the transverse wave vector is linear rather than exponential as for the case of LHM \cite{Pendry}.

In the above analysis, we have assumed the wire medium to be a homogeneous dielectric material. However in practice, since the electric displacement inside the wire medium is proportional to the currents flowing along the wires, in order to detect the internal images, one needs to measure the current density instead of either electric or magnetic field, in contrast to the canalisation regime that the image is formed at the back interface of the wire medium and can be detected using the near-field scanning. In the following section, we model the physical structure of the wire medium and demonstrate numerically that the currents along the wires indeed carry the subwavelength details of a source.

\section{Simulations of internal imaging by modelling the physical structure of the wire medium}
\begin{figure}[htbp]
\centering
\includegraphics[width=13cm]{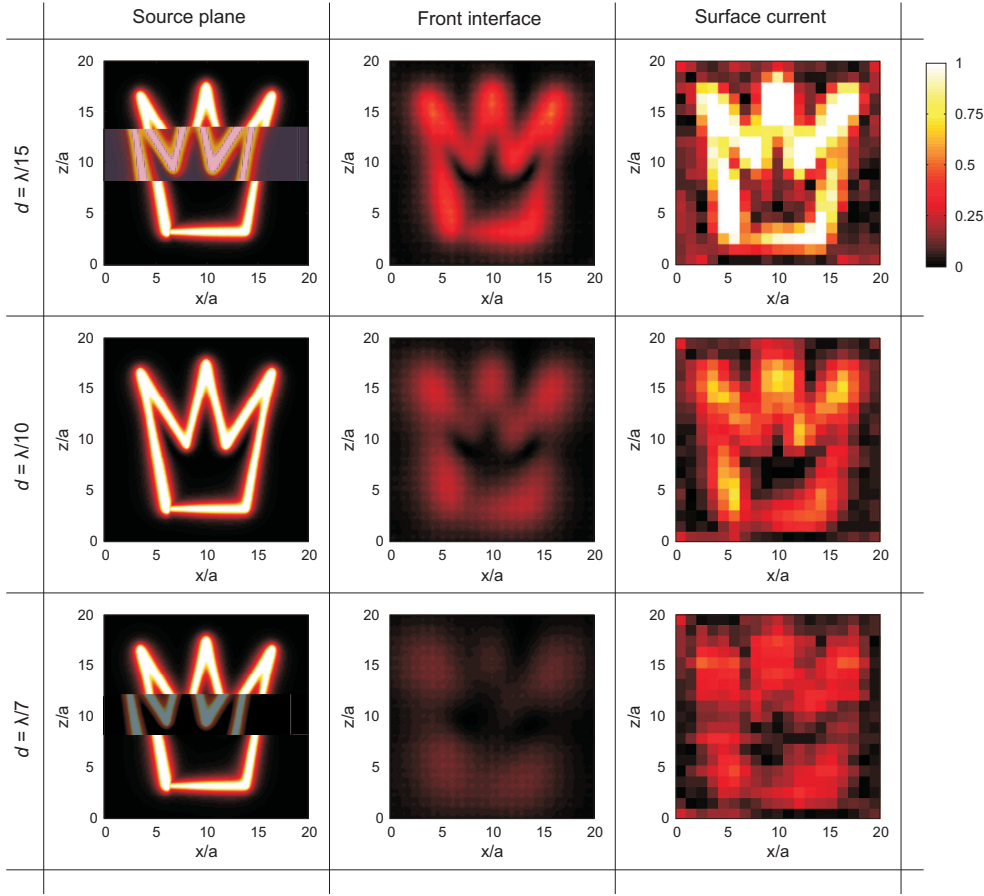}
\caption{Normalized distributions of the $y$-component of the electric field at the source plane, the front interface, and the current density at the internal image plane of the wire medium for different distances between the source and the wire medium calculated from CST simulations of the physical structure.}
\label{fig_cst}
\end{figure}

In the simulations using CST Microwave Studio$^{\rm TM}$, the overall dimensions of the wire medium and the crown-shaped source are kept the same as in the previous section using the FDTD method. The number of wires is $21\times21$ along $x$- and $z$-directions respectively, and the inner spacing between wires is $a=\lambda/30$. The radius of the wires is $r=\lambda/300$. We also consider three different distances between the source and the wire medium: $d=\lambda/15$, $d=\lambda/10$ and $d=\lambda/7$. The normalized distributions of the $y$-component of the electric field at the source plane, at the front interface, as well as the current density along the wires (the currents are converted to field intensities to compare with the other two distributions) in the middle plane of the wire medium are shown in Fig.~\ref{fig_cst}.
It is shown that the source distribution is much sharper compared to the FDTD simulation results. This is due to the fine mesh used in CST simulations in order to model thin wires accurately. Again we observe similar distributions at the front interface of the wire medium where less details of the source can be detected. The current density is plotted on the discrete grids (the third column in Fig.~\ref{fig_cst}) since only $21\times21$ wires are present in the modelled wire medium. The current density shows significant enhancement compared to the field distributions at the front interface of the wire medium, which demonstrates the internal subwavelength imaging capability of the wire medium.

The low image resolution of the internal current density distribution is due to the insufficient sampling points ($21\times21$). Therefore to acquire a higher resolution of internal images, one can use the wire medium formed by more densely packed wires. For example, we have considered a wire medium formed by $41\times41$ wires with the spacing between the wires $a=\lambda/60$, keeping the transverse dimension of the wire medium the same as the previous case. According to Eq. (\ref{eq_k0}), the plasma frequency is higher for this case and therefore one could expect images with a higher resolution. However the wire density cannot be increased arbitrarily due to the fact that when the filling ratio of wires increases, the structure may exhibit anisotropic magnetic properties \cite{Tyo,Hu,Shin}. The influence of the material parameter change on the subwavelength imaging property of the wire medium is currently under our investigation.

\begin{figure}[htbp]
\centering
\includegraphics[width=14cm]{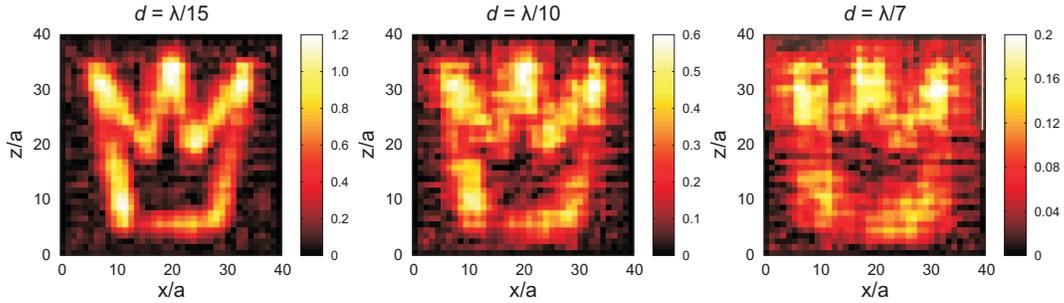}
\caption{The current density from CST simulations of the physical wire medium ($41\times41$ wires) for different distances between the source and the wire medium.}
\label{fig_cst2}
\end{figure}

The current density distributions from CST simulations for the case of $41\times41$ wires are plotted in Fig.~\ref{fig_cst2}.
Note that for illustration purposes, the distributions are not normalized. It can be seen that by increasing the density of wires, the internal image resolution can be improved, which is especially useful when a more complex source is considered.

The preliminary estimations show that the image appears near the middle plane of the $\lambda/2$ thick wire medium slab and does not change significantly even when the image plane is offset from the middle plane by 5-10\%. The plane of the best image formation cannot be determined precisely from numerical simulations since the distribution of electric displacement or current density inside the wire medium varies slowly along the direction of wires. We have considered a wire medium slab with a half-wavelength thickness in this paper. For the wire medium slabs with greater thicknesses (an integer multiple of a half-wavelength), there are multiple planes for the internal imaging, where the evanescent spatial harmonics generated by the source can be amplified and the information lost in the free space region can be restored. The number of internal image planes can be given by $N$ if the thickness of the wire medium is $N\lambda/2$. The distance of the internal image planes to the front interface of the wire medium can be approximately given by $(2n-1)\lambda/4$ where $n=1,2,...,N$.

\section{Conclusion}
The internal imaging capability of the wire medium, i.e. images with a much higher resolution appear inside of the wire medium comparing with the external ones, is demonstrated numerically by modelling the physical structure and using the effective medium theory: the wire medium slabs are capable of imaging a source with subwavelength resolutions located at significant distances away from the device. The operational principle of `internal imaging' is based on the excitation of standing waves inside the wire medium slab rather than the resonance of surface plasmons in LHMs. The features of being low loss and easy in manufacturing make the wire medium attractive for practical applications. In practice, it is possible to capture the internal image by embedding detectors into metallic wires at the internal image plane and directly measuring the currents flowing along the wires. The internal imaging effect can be used for subsurface imaging and the creation of new generation subwavelength imaging devices.

\section*{Acknowledgements}
The authors would like to thank EPSRC for the financial support and Mr. G. Palikaras for his help during the preparation of the manuscript. P. A. Belov acknowledges the financial support by EPSRC Advanced Research Fellowship EP/E053025/1.


\section*{References}
\begin{thebibliography}{99}

\bibitem{Pendry}
J. Pendry, ``Negative refraction index makes a perfect lens,'' Phys. Rev. Lett. \textbf{85}, 3966-3969 (2000).

\bibitem{Veselago}
V. G. Veselago, ``The electrodynamics of substances with simultaneously negative values of $\varepsilon$ and $\mu$,'' Sov. Phys. Usp. \textbf{10}, 509-514 (1968).

\bibitem{Podolskiy}
V. A. Podolskiy, and E. E. Narimanov, ``Near-sighted superlens,'' Opt. Lett. \textbf{30}, 75-77 (2005).

\bibitem{Smith}
D. R. Smith, D. Schurig, M. Rosenbluth, S. Schultz, S. A. Ramakrishna, and J. B. Pendry, ``Limitations on subdiffraction imaging with a negative refractive index slab,'' Appl. Phys. Lett. \textbf{82}, 1506 (2003).

\bibitem{BelovCanal}
P. A. Belov, C. R. Simovski, and P. Ikonen, ``canalisation of subwavelength images by electromagnetic crystals,'' Phys. Rev. B \textbf{71}, 193105 (2005).

\bibitem{BelovLayered}
P. A. Belov and Y. Hao, ``Subwavelength imaging at optical frequencies using a transmission device formed by a periodic layered metal-dielectric structure operating in the canalisation regime,'' Phys. Rev. B \textbf{73}, 113110 (2006).

\bibitem{Wood}
B. Wood, J. B. Pendry, and D. P. Tsai, ``Directed subwavelength imaging using a layered metal-dielectric system,'' Phys. Rev. B \textbf{74}, 115116 (2006).

\bibitem{Scalora}
M. Scalora \textit{et al.}, ``Negative refraction and sub-wavelength focusing in the visible range using transparent metallodielectric stacks,'' Opt. Express \textbf{15}, 508-523 (2007).

\bibitem{Liu}
Y. Liu, G. Bartal, D. A. Genov, and X. Zhang, ``Subwavelength discrete solitons in nonlinear metamaterials,'' Phys. Rev. Lett. \textbf{99}, 153901 (2007).

\bibitem{BelovWire}
P. A. Belov, Y. Hao, and S. Sudhakaran, ``Subwavelength microwave imaging using an array of parallel conducting wires as a lens,'' Phys. Rev. B \textbf{73}, 033108 (2006).

\bibitem{ZhaoOE}
Y. Zhao, P. A. Belov, and Y. Hao, ``Spatially dispersive finite-difference time-domain analysis of sub-wavelength imaging by the wire medium slabs,'' Opt. Express \textbf{14}, 5154-5167 (2006).

\bibitem{BelovSpatial}
P. A. Belov, R. Marques, S. I. Maslovski, I. S. Nefedov, M. Silverinha, C. R. Simovski, and S. A. Tretyakov, ``Strong spatial dispersion in wire media in the very large wavelength limit,'' Phys. Rev. B \textbf{67}, 113103 (2003).

\bibitem{WMJEWA}
P. A. Belov, S. A. Tretyakov, and A. J. Viitanen, ``Dispersion and reflection properties of artificial media formed by regular lattices of ideally conducting wires,'' J. Electromagn. Waves Applic. \textbf{16}, 1153-1170 (2002).

\bibitem{Taflove}
A. Taflove, Computational electrodynamics: the finite-difference time-domain method, 2nd ed., Artech House, Norwood, MA (2000).

\bibitem{Berenger}
J. R. Berenger, ``A perfectly matched layer for the absorption of electromagnetic waves,'' J. Computat. Phys. \textbf{114}, 185-200 (1994).

\bibitem{ZhaoAP}
Y. Zhao, P. A. Belov, and Y. Hao, ``modelling of wave propagation in wire media using spatially dispersive finite-difference time-domain method: numerical aspects,'' IEEE Trans. Antennas Propagat. \textbf{55}, 3070-3077 (2007).

\bibitem{Tyo}
J. S. Tyo, ``A class of artificial materials isorefractive with free space,'' IEEE Trans. Antennas and Propagat. \textbf{51}, 1093 (2003).

\bibitem{Hu}
X. Hu, C. T. Chan, J. Zi, M. Li, and K.-M. Ho, ``Diamagnetic response of metallic photonic crystals at infrared and visible frequencies,'' Phys. Rev. Lett. \textbf{99}, 223901 (2006).

\bibitem{Shin}
J. Shin, J.-T. Shen, P. B. Catrysse, and S. Fan, ``Cut-through metal slit array as an anisotropic metamaterial film,'' IEEE J. Selected Topics Quantum Electronics \textbf{12}, 1116-1122 (2006).

\end{thebibliography}
\end{document}